\documentclass[useAMS,usenatbib]{mn2e}

\usepackage{amssymb,amsmath}
\usepackage{amsfonts} 
\usepackage{psfig}
\usepackage{graphicx}
\newcommand{ \cgs}{erg cm$^{-2}$ s$^{-1}$}
\newcommand{ \psu}{cts$^{2}$ s$^{-2}$ deg$^{-4}$}

\title[The nature of the unresolved  extragalactic  cosmic soft X-ray background]{The nature of the unresolved extragalactic soft CXB}
\author[N. Cappelluti et al.]{N. Cappelluti$^{1,2}$\thanks{E-mail:
nico.cappelluti@oabo.inaf.it} ,
P. Ranalli$^{3,4,1}$, M. Roncarelli$^{4}$, P. Arevalo$^{5}$, G. Zamorani$^{1}$ 
\newauthor
A. Comastri$^{1}$, R. Gilli$^{1}$, E. Rovilos$^{1}$, 
C. Vignali$^{4,1}$, V. Allevato$^{6}$, A. Finoguenov$^{2,7}$, 
\newauthor
T. Miyaji$^{8}$, 
F. Nicastro$^{9}$,  I. Georgantopoulos$^{3,1}$ and A. Kashlinsky$^{10,11}$ \\
$^{1}$INAF-Osservatorio Astronomico di Bologna, Via Ranzani 1, 40127 Bologna, Italy\\
$^{2}$University of Maryland, Baltimore County, 1000 Hilltop Circle, Baltimore, MD 21250, USA\\
$^{3}$ Institute of Astronomy and Astrophysics, National Observatory of
    Athens, Palaia Penteli, 15236 Athens, Greece\\
$^{4}$ Universit\`a di Bologna, Dipartimento di Astronomia,
    via Ranzani 1, 40127 Bologna, Italy \\
$^{5}$Departamento de Ciencias Fisicas, Universidad Andres Bello, Av.
Republica 252, Santiago, Chile\\
$^{6}$Max-Planck-Institut f\"{u}r Plasmaphysik and Excellence Cluster Universe, Boltzmannstrasse 2, D-85748 Garching, Germany\\
$^{7}$Department of Physics, University of Helsinki, Gustaf HŠllstršmin katu
2a, FI-00014 Helsinki, Finland\\
$^{8}$Instituto de Astronom'a, Universidad Nacional Autonoma de Mexico, Ensenada, Baja California, \\
  ~~Mexico (mailing address: P.O. Box 439027, San Diego, CA 92143-9027, USA)\\
$^{9}$Osservatorio Astronomico di Roma (INAF), Via Frascati 33, I-00040 Monte Porzio Catone, Italy\\
$^{10}$SSAI, Lanham MD 20706, USA\\
$^{11}$ Observational Cosmology Laboratory, Code 665, Goddard Space Flight Center, Greenbelt, MD 20771, USA\\
  }
\begin{document}

\date{}

\pagerange{\pageref{firstpage}--\pageref{lastpage}} \pubyear{2002}

\maketitle

\label{firstpage}

\begin{abstract}
In this paper we investigate the power spectrum of the unresolved  0.5-2 keV CXB
with deep {\em Chandra} 4 Ms  observations in the CDFS.
We measured a 
signal which, on scales $>$30$\arcsec$, is significantly higher than the Shot-Noise
and  is increasing with the  angular scale.  
We interpreted this signal as the joint contribution of clustered undetected sources like AGN, Galaxies
and Inter-Galactic-Medium (IGM).   The power of unresolved cosmic sources 
fluctuations accounts for  $\sim$12\% of  the 
 0.5-2 keV extragalactic CXB.  
Overall,  our modeling predicts that $\sim$20\% of the unresolved CXB flux is made 
by low luminosity AGN, $\sim$25\% by galaxies and $\sim$55\% by the IGM (Inter Galactic Medium).
 We do not find any direct evidence of the so called Warm Hot Intergalactic Medium 
(i.e. matter with  10$^5$K$<$T$<$10$^7$K and 
density contrast $\delta<$1000),
but we estimated that it could produce about 1/7  of the unresolved CXB.   

We placed an upper limit to the space density of postulated X-ray-emitting early black hole at 
 z$>$7.5  and compared it  with 
SMBH evolution models. 
\end{abstract}


\section{Introduction}
 The Cosmic X-ray Background (CXB) is the 
 result of a multitude of energetic phenomena 
 occurring in the Universe since the epoch of the formation of 
the  first galaxies.  Its nature has been investigated in the last 50 
 years with several telescopes but only in the 80's  it became clear that 
 its main contributors are AGN \citep{suaeccellenza}. 
 Later it has been found that also  galaxies, galaxy clusters, large scale structures 
and  diffuse hot gas in the Milky Way are sources contributing to the CXB \citep{fb}. \\
With the launch of ROSAT, {\em Chandra} and XMM-{\em Newton},  the knowledge
about the nature of  sources contributing to the flux of the CXB
became suddenly clear.  Deep surveys like the Chandra Deep Field South \citep[CDFS,][]{gia,luo,xue}
and North \citep[CDFN,][]{bra}, and the Lockman Hole \citep{brun}  have almost conclusively resolved 
 the problem of the  CXB below 10 keV. In fact  at the flux limits of  {\em Chandra} and XMM-{\em Newton},
 about 90-95$\%$ \citep{bauer,moretti,lehmer12} of the 0.5-2 keV CXB flux has been resolved\footnote{In these papers,
 the fraction of unresolved CXB has been estimated at the flux of the faintest 
 detected source, here we measure the average value on  the investigated area} into point and extended sources. \\
 CXB synthesis models \citep[see e.g.][]{tre06,gil07}
predict that at   fluxes fainter that the current limits, the AGN  source counts progressively
flatten and then galaxies become the more abundant sources.\\
 Galaxies emerge as the dominant population at faint 0.5-2 keV
fluxes,
with break-even  point at around $\sim$10$^{-17}$ \cgs \citep[see also][]{lehmer,xue,lehmer12} 
and can make up for a sizable fraction of
the X-ray background (9-16\%; Ranalli et al. 2005). 
The contribution from galaxy clusters, down to a mass 
limit of $\sim$10$^{13}$ M$_{\odot}$, has been estimated to be 
of the order of $\sim$10\% of the total 0.5-2 keV  CXB \citep{gil99,lemze}. \\
  Several authors studied the clustering properties of the CXB
 to unveil the nature of the sources producing its flux and their properties
 \citep{mart,wu,sol94,bf}. The most used technique is that of the two-point 
 autocorrelation function of the surface brightness of the CXB. 
 While at the time of HEAO-1 the task
 was to determine what was the contribution of QSO to the CXB, after ROSAT
 most of the  investigations in this field have been performed to unveil signatures of the 
 Warm Hot Intergalactic Medium \citep{sol02,sol06,gale} and to test Cosmological models \citep{diego}.  
 This kind of studies is  therefore extremely powerful to study population of
 sources which are beyond the resolving and detection capabilities of instruments.

In the GOODS fields, \citet{HM06, HM07} have shown that, after excising detected point and extended sources plus
 faint HST detected galaxies, the spectrum of the soft CXB was still showing a signal in 
the 0.5-2 keV band. While a large amount of it could be attributed to a local component 
 (Solar Wind Charge Exchange and Milky Way thermal emission), they have shown that 
below 1 keV  the fraction of the resolved CXB is not sensitive to the 
 removal of HST galaxies,  thus arguing for a purely diffuse nature of the 
 unresolved CXB.  In the 1-2 keV band they have shown that, by  removing HST sources 
 areas from the analysis, the fraction of unresolved CXB drops by $\sim$50\%.
 This means that a large fraction of the unresolved 1-2 keV CXB flux could be produced
 in faint undetected galaxies (active or non-active). 
  Moreover, they speculated that a large fraction of 
 the unresolved CXB may be due to faint 3.6$\mu$m IRAC sources, suggesting 
that  high-z or absorbed sources could produce a large part of such
a radiation. 
 According to their estimate, the remaining fraction of the CXB (i.e. 2-3\% of the 0.5-2 keV CXB)
 remains consistent with the prediction of Warm Hot Intracluster Medium (WHIM) intensity from 
 hydrodynamical simulations \citep[see e.g][]{cen,ursino,mauro06,ron}.\\
In the local Universe about 30-40\% of  the baryons are missing with respect to what is measured
at z$\sim$3 \citep{fuku}. Simulations predict that most of these baryons got shock-heated and at z$\lesssim$1 
they should have a temperature of the order 10$^{5}$-10$^{7}$ K and therefore emit thermal
X-rays \citep{cen}.   Controversial evidences on the properties of such 
a medium have been published so far \citep[see e.g.,][]{nicastro,kaastra,fang,shull}.
Though with  very low surface brightness, the WHIM can be distinguished from other 
kinds of diffuse emissions on the basis of  its clustering properties \citep{ursino}.  In fact, its emission
follows that of clusters and filaments and peaks at  low redshift (z$\sim$0.5), thus 
showing a typical feature in the angular clustering  of unresolved  CXB fluctuations.
However, WHIM is not the only expected component of the unresolved CXB
arising from thermal emission of the Inter Galactic Medium (IGM). 
X-ray surveys in the local Universe revealed X-ray emission from local galaxy groups 
down to masses of the order of 10$^{12}$ M$_\odot$ \citep[see e.g.][]{eck}.
Since the intensity of the X-ray emission of galaxy groups scales
with their mass, a large amount of them has not been detected
at moderately high-z. Therefore we also expect a contribution to the overall
signal from medium to low mass-groups  at z$>$0.2-0.3.\\
Concerning the unresolved extragalactic CXB, it is difficult to distinguish its 
components with a simple spectral analysis mostly because of the poor 
energy resolution of X-ray sensitive CCDs (i.e. $\sim$130 eV, for ACIS-I at 1.5 keV). 
However, cosmological sources leave a unique imprint in the power spectrum (PS) 
of the anisotropies of the fluctuations of the unresolved CXB in a way that is related
to their clustering and volume emissivity properties. \\
The unresolved CXB contains information about all those sources that have 
not been detected at the deepest  fluxes reachable by deep surveys.
Moreover, the amplitude of the PS is not only sensitive to the luminosity density of those
sources, but it also provides information about their  bias.\\

In this paper we study the power spectrum of the 
unresolved 0.5-2 keV CXB with the  4Ms CDFS data,
the deepest X-ray observation ever performed to date.  We model the anisotropies of unresolved 
CXB with the state-of-the art results on galaxy and AGN evolution models and observations
as well as with modern hydrodynamical cosmological  simulations. \\
Throughout this paper we will adopt a concordance $\Lambda$-CDM cosmology with $\Omega_m$=0.3,
 $\Omega_{\Lambda}$=0.7, H$_0$=70 h km/s/Mpc and $\sigma_8$=0.83. 
 Unless otherwise stated, errors are quoted at 1$\sigma$ level and fluxes refer to the 0.5--2 keV band.
We used as reference cumulative 0.5-2 keV CXB flux the recent estimates of \citet{lehmer12},
S$_{CXB}$(0.5-2)=8.15$\pm{0.58}\times$10$^{-12}$ \cgs~deg$^{-2}$.

\section{Dataset}
\begin{figure*}
   \centering
\hspace{1cm}
 \includegraphics[width=0.85\textwidth]{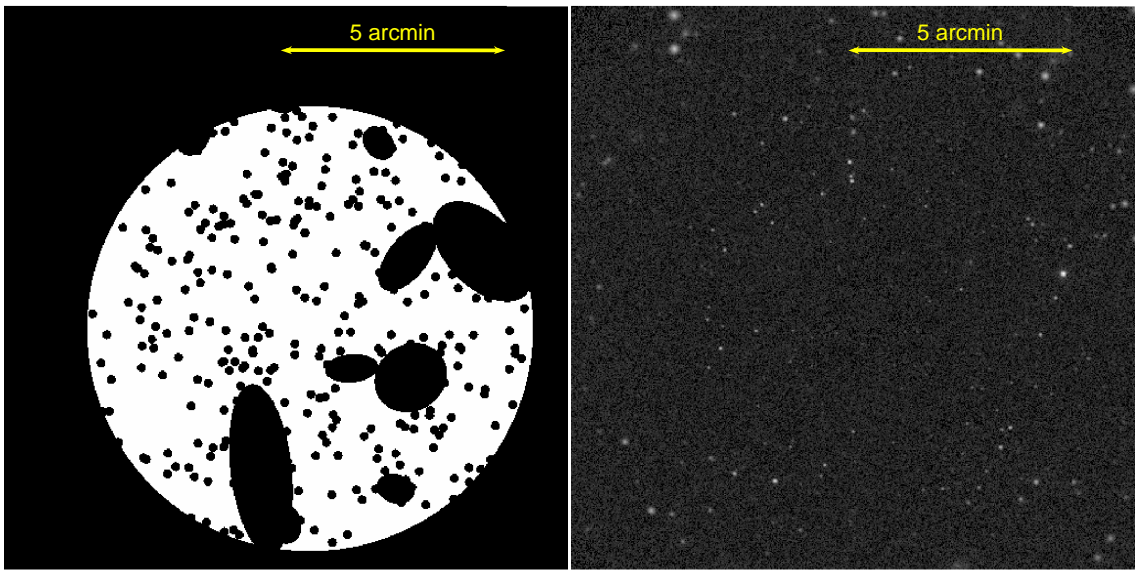}
   \caption{ $Left:$The mask adopted for the ACF analysis in the  observations of the CDFS. 
	    The white area is the resulting source-free area. $Right:$ 0.5-2 keV image of the same area smoothed with a Gaussian
	    filter with 3$\arcsec$ width.}
              \label{mask}
    \end{figure*}

For the purposes of this work we used
 the 4 Ms Chandra exposure on the  CDFS \citep{xue}.
 This is the  deepest Chandra observation ever performed and reaches 
 a point source flux limit of $\sim$10$^{-17}$  \cgs. A detailed description of the 
 data reduction and catalogue production is given in \citet{xue}. 
In order to improve the sensitivity on extended sources, we also used 
the  3 Ms XMM-Newton exposure in the same field \citep{piero}.
Since the scope of this paper is to measure the angular PS  (see next section) of the purely diffuse, unresolved CXB, we  
masked the  Chandra data from all the point sources detected by \citet{xue}.
The combination of the Chandra and XMM-Newton observations allowed us to remove, 
with unprecedented sensitivity, also extended
sources down to 0.5-2 keV fluxes of 2$\times$10$^{-16}$  \cgs~ (Finoguenov et al. , private communication). 
In order  to employ the best-PSF area, we limited 
our analysis to the inner 5$\arcmin$ from the exposure weighted mean 
optical axis ($\alpha$=03:32:27.316, $\delta$=-27:48:50.36). Point sources have been masked with   circular regions
of 5$\arcmin\arcmin$ radius (which is large enough to mask $>$99\% of the source
flux at every off-axis angle investigated here) and extended sources have been masked out to R$_{200}$ (i.e. radius
within which the matter overdensity is $>$200). In this way the remaining counts on the detector are  made, 
with  good approximation,  by particle background and purely
diffuse cosmic background only.
 With such a masking our image  contains 163940 counts, with an average 
occupation number of 1.7 cts/pix. 
 The total investigated masked area  (80.3 arcmin$^{2}$)  
is  shown in Fig. \ref{mask} together with  the raw image 
of the same region. The source masking removes a fraction of the 
useful  area of about $\sim$25\% and therefore the source-free area 
is of the order of 60 arcmin$^{2}$.     In principle the evaluation of the PS should not
change if we increase the size of the mask. However by increasing  the mask size
we would run into an over-masking problem that would bias the estimate of the PS. 
If the masking were larger than $\sim$30\%-40\%
of the active pixels then the PS analysis would have been biased by the mask \citep{kash12}.
A smaller size of the mask would increase the overall power since it would 
be polluted by power from cluster outskirts and  PSF tails of detected sources.
In order to improve the signal to noise ratio of the weak  signal we are looking for, 
we applied a careful filtering of the background.
We have additionally   filtered the events files provided by the CXC  from particle flares
by using the ciao routine $lc\_clean$, which performs a 3$\sigma$ clipping of the 
background light-curve\footnote{http://cxc.harvard.edu/ciao/ahelp/lc$\_$clean.html}. Such a technique  is more 
sensitive than the classical  procedures adopted in standard pipelines.  
In order to reduce the noise introduced by low-count statistics,
 images were binned in pixels of 1.5$\arcsec$ size.
 Images have been produced in the 0.5-2 keV energy range, where {\em Chandra} has the
 largest collecting area and detector efficiency.
\section{Evaluation of the power spectrum}
 \label{tech} 
In order to estimate the power spectrum we adopt the method described in 
\citet{are} and \citet{chura} that has been developed with the purpose of estimating the PS
of 2D images with gaps.   Here we give a brief description of the technique used to evaluate
the PS, but we refer the reader to \citet{are} for a more comprehensive description.\\
The first step is to derive the fluctuation field at any given angular scale $\theta$,
from which we derive a low-resolution PS by  using a Mexican-Hat (MH) filter, 
$F(x)=[1-\frac{x^2}{\theta^2}]e^{-\frac{x^2}{2\theta^2}}$.\\
The goal is to  single out fluctuations at any given angular scale and then  compute the variance
of the fluctuation field image, $\delta_S^{2}(k)$.
 Our map has gaps introduced by features in the exposure maps 
and by the mask. The mask (M) is an image having value M=0  
in the region of excised sources and M=1 elsewhere. 
 The method corrects  for gaps by representing the MH filter
as a difference  between two Gaussian filters with slightly different widths, convolving the image 
and the mask with these filters and dividing the results before calculating the final filtered images.
It has been shown \citep{are} with numerical simulations that this method efficiently takes into account 
data masking also when a large fraction of the field is masked and, independently from the mask shape,
the normalization and the shape of the PS is accurately estimated. 
 At any given frequency  $k$=1/$\theta$, we estimated the PS of the fluctuation field as follows:

 We define the fluctuation  image at the scale $\theta$ as:\\
\begin{equation}
I_{\theta}=M \times \left[\frac{G_{\theta_1} \circ (I)}{G_{\theta_1} \circ M}-\frac{G_{\theta_2} \circ (I)}{G_{\theta_2} \circ M}\right].
\label{est}
\end{equation}
where (G$_{\theta_1} \circ$ I, G$_{\theta_2} \circ$ I) represent the count-rate image (I)
convolved with a gaussian filters with widths $\theta_{1,2}$ (G) and M is the mask\footnote{The symbol $\circ$ stands for 
convolution}. 
The width of the filter is chosen as $\theta_{1}=\theta\sqrt{1+\epsilon}$ and 
$\theta_{2}=\theta/\sqrt{1+\epsilon}$, where $\epsilon<<$1.
In this way the value in the brackets of eq.\ref{est} is dominated by fluctuations at the scale $\sim\theta$. \\
\citet{are} demonstrated that PS of the fluctuation field at the 
 frequency $k$=1/$\theta$  (P$_2$(k))  is related to the variance of image via:
\begin{equation}
P_2(k)=\frac{1}{\epsilon^2\pi k^2}\sigma^{2}(I_{\theta})=<|\delta_S(k)|^2>.
\label{eq:estim}
\end{equation}
In this work we used  $\epsilon=$0.1,  however a different value of $\epsilon$ 
does not change the estimate of the PS, provided that  $\epsilon<<$1.
Note that in the calculation of the variance we have made no assumption on the
distribution of the fluctuation amplitude, thus the method works also in the Poisson 
regime.
 This method is equivalent to the classical Fourier analysis;  the only 
difference is that this treatment of the gaps and  choice of the filters 
allow us to retrieve the underlying  power-spectrum regardless of the shape
of the mask.
Unless otherwise stated, errors are given at the $1\sigma$ level. 
  In Fourier analysis, the uncertainty on power estimate is 
 solely related to the accuracy with which one can estimate the variance. 
 At every frequency, this depends on the number of independent elements
 in the Fourier space, and the intrinsic dispersion of the data. 
 In our analysis, since we adopt a broad filter, we need to define, at every frequency $k$,
 the mean number of real Fourier elements used to evaluate the amplitude of the fluctuations
 $<N_k>$. In principle the uncertainties  can be approximated with $\sigma_P(k) \approx \,P_2(k)/\sqrt{0.5<N_k>}$.
Following the extensive description of \citet{are},
within our formulation errors are   computed with:
\begin{equation}
 \sigma_P(k)=\frac{\sqrt{\int_q(P(k)\,F_{qx}^2dq}}{F^2_{kx}}\frac{N}{N_{M=1}},
\end{equation}
where F$_{kx}$ is the Fourier transform of the filter, N is total number of pixels and
N$_{M=1}$ is the number of unmasked pixels.\\
   This filter-weigthed   expected uncertainties
   is what we plot as error bars. 

Power spectrum points measured in this way
 are partially dependent on their neighbor frequencies as a result of 
 the finite width of the power filter. 
 Therefore, points can only be considered independent 
 if they are measured at spatial scales that are sufficiently separated. 
 In our case, where we used a factor of 2 separation in angular scale, it  is appropriate to use PS 
 estimates and errors as independent points in the usual fitting methods.
\section{Results}

\subsection{Baseline method}

\begin{figure}
   \centering
 \includegraphics[width=0.45\textwidth]{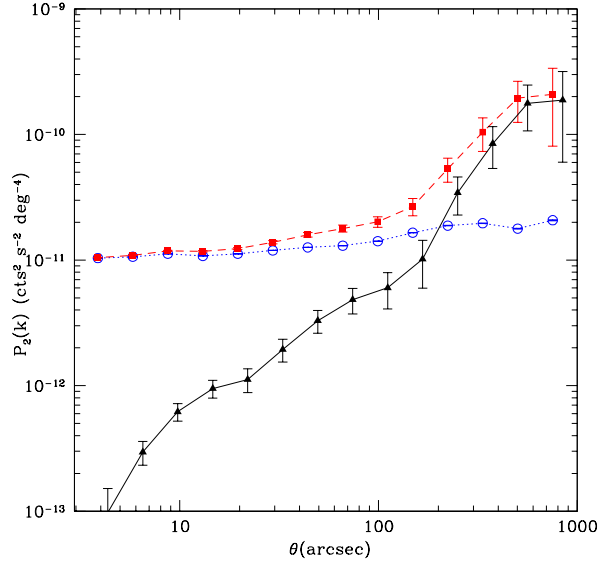}
   \caption{ $Red~squares:$ the raw PS of the source-masked fluctuations of the CXB as a function of the angular scale. $Blue~open~circles:$ the (A-B)/2
   PS.  $Black~Triangles:$ The resulting Cosmic PS
   after subtracting the "spurious" power. $Black~symbols$ are artificially shifted for sake of clarity.}
              \label{rawps}
    \end{figure}
The power spectrum estimated with the technique described is Sect. \ref{tech} is shown in Fig. \ref{rawps}. 
We computed the probability that the observed data could be 
produced  by a statistical fluctuation of a zero power signal by using a simple
$\chi^2$ test. We obtained a value of $\chi^2$=140 corresponding 
to  a $>$10$\sigma$ significance with respect to zero power.
The high frequency part is dominated by a white noise component, while
 the signal increases towards lower frequencies (i.e. large scales). In {\em Chandra}  images
 a large fraction of the fluctuation can be attributed to Poisson noise due to the low-count
 statistics.

 We estimated $P_{noise}$ directly from the 
 data by using the A-B technique \citep{kash05}. 
 For every pointing we sorted the events  in order of time arrival 
 and created two images, one including even events (A), and the other
 with odd events (B). We have then mosaiced all the A and B images.
 In this way we have two identical images with half the photons with respect 
 to the full dataset, having both the same exposure, thus containing 
 the same information on the signal but with their own noise. 
 Any systematic errors should be manifested very similarly in the A 
 and B images, and thus the (A-B)/2 difference images provide
  a useful means of characterizing the random (noise) properties of the data sets.
 The  PS of the difference  image (A-B)/2 is therefore nothing else than
 the PS of fluctuations due to Poisson noise. Source variability does not affect this procedure
 since the sampling  of their light curve does not change significantly:  the gap between 
 two consecutive events is of the order of the read-out time of the detector, which is much shorter 
 than the typical variability of X-ray sources. 
In Fig. \ref{rawps} we show the PS of the Poisson noise compared with the PS
of the raw image. As expected, the Poisson Noise shows a flat  (white noise) spectrum.
Throughout this paper we assume that both particle and galactic background
contribute only to the Poisson noise since their flux is expected to be homogeneous 
at the scales sampled by this investigation.  
The total cosmic PS (P$_{2,CXB}$) is therefore obtained with:
\begin{equation}
P_{2,CXB}=P_{2,tot}-P_{2,(A-B)/2}.
\end{equation}
The estimated P$_{2,CXB}$ is plotted in Fig. \ref{rawps}. 
In the angular range sample here (3$\arcsec<\theta<1000\arcsec$), the power
increases with the scale, getting close to zero at smallest scale.
 The errors on P$_{2,CXB}$ have been obtained by propagating the errors of P$_{2,tot}$ and  P$_{2,(A-B)/2}$.
\section{Modeling the CXB anisotropies}
 In order  to interpret the cosmic signal measured in the 
PS of the CXB, we used population synthesis models and 
recent observational results to predict amplitude and 
shape  of its components.
 We want to test the assumption that the total observed fluctuations
can be reproduced by a simple 3 source class model.  
 Our hypothesis is that the amplitude and the shape of the  PS are produced by:
Shot Noise from individual undetected AGN and galaxies within the instrument
 beam (discrete source counts),  clustering of galaxies and 
AGN below the limiting flux, clustering of undetected diffuse
 hot cosmological gas in large scale structures (IGM, i.e. undetected groups
and WHIM  filaments).  We then test the statistical robustness of our hypothesis with 
$\chi^2$ test where every component is considered as a   parameter. In addition we provide an
upper limit for the still undetected high-z sources.
We assumed that, on  the angular scales investigated here,  fluctuations from the Milky Way diffuse 
emission does not contribute to the PS,  i.e. this emission can be described  as a constant flux on the 
detector and therefore contributing to the Poisson noise only. This hypothesis is supported by the PS
measurements of the soft CXB with ROSAT \citep{silwa}  
that shows that the Galaxy produces signal only for the smallest harmonics
of the  PS. Moreover, cosmological sources (AGN, Galaxies and IGM) partially share the same environment, thus 
we include in the model a cross power term to model the amplification of fluctuations produced by their cross-correlation.    
  In summary, the CXB PS can be modeled as:
\begin{equation}\begin{split}
P&_{2,CXB}(k)=P_{2,SN}(k)+P_{2,AGN}(k)+\\
&P_{2,GAL}(k)+P_{2, IGM}(k)+P_{2,A,G,I}(k),
\label{hyp}
\end{split}\end{equation}
where $P_{2,SN}(k), P_{2,AGN}(k), P_{2,GAL}(k)$ and $P_{2, IGM}(k)$
are the PS of shot noise and the clustering components of  AGN, Galaxies
and IGM, respectively.  P$_{2,A,G,I}(k)$ is the cross power term
which contains the cross-correlation of AGN and galaxies, AGN and IGM,
Galaxies and AGN, respectively. 
 In the following sections  we will describe 
the procedure adopted to model each spectral component.

\subsection{Preliminaries}

To decompose  P$_{2,CXB}$ in its primary components
we adopt the procedure described by \citet{kash}.
Whenever a fraction of the sky of the order $\theta<$1 rad is concerned, one adopts 
the Cartesian formulation of the Fourier analysis. 
The fluctuation field of the CXB surface brightness (S) is defined as $\delta$S=S$(\theta)-\langle S \rangle$. 
 P$_2(k)$ is related to the auto-correlation function  C($\theta$) through: 
 \begin{equation}
 P_2(k)=\int_0^\infty C(\theta)J_0(k\theta)kdk,
\end{equation}
where J$_0(x)$ is the zero$-th$ order cylindrical Bessel function. 
A very useful quantity is the mean square fluctuation within a finite beam of angular radius
$\theta$, or zero-lag correlation, which is related to the PS via:
\begin{equation}\begin{split}
&C(0)=\langle \delta S^2 \rangle_{\theta}= \\
&\frac{1}{2\pi}\int_0^{\infty}P_2(k)W_{TH}(k\theta)k dk  \propto k^2P_2(k),
\end{split}\end{equation}
where W$_{TH}$ is the window function. 
Thus, in order to visualize the relative strength of the fluctuations at any given scale 1/$k$, 
a useful quantity is $\sqrt{k^2\,P_2(k)}$, that differs from the actual value
of $<\delta_{S}>_{\theta}$ by a  factor of the order unity that depends on the window function.
\subsection{Shot noise level and PSF modeling}
The shot noise  is  produced by discrete sources in the
beam. The relative amplitude of this component scales as N$^{-1/2}$, where N is the mean number 
of sources entering the beam. P$_{2,SN}$ is therefore 
very important in surveys performed  by instruments with   high angular resolutions like
{\em Chandra}. If sources are removed down to a flux S$_{lim}$, it can be shown \citep{kash} that the shot noise component 
can be expressed as:
\begin{equation}
P_{2,SN}=\int^{S_{lim}}_{0}S^{2}\frac{dN_X}{dS}dS,
\label{sn}
\end{equation}
where $\frac{dN_X}{dS}$ is the differential logN-logS of all the X-ray point  sources
 (i.e. AGN, Galaxies, Stars).  
In most cases, the sensitivity of the survey is not homogeneous 
 across the field of view. This means that S$_{lim}$
 in Eq. \ref{sn}  is a function of the sky coordinates. 
 To account for  this selection effect we scaled the 
 number counts by the selection function computed as follows:§
 we computed the sensitivity map of the surveyed area 
 by imposing the same false source detection rate  adopted 
 by \citet{xue} for catalogue production (i.e. P$_{false}<$0.004). 
By using the masked maps described above we estimated the minimum 
count rate necessary not  to  exceed the threshold mentioned above. 
\begin{figure}
   \centering
 \includegraphics[width=0.45\textwidth]{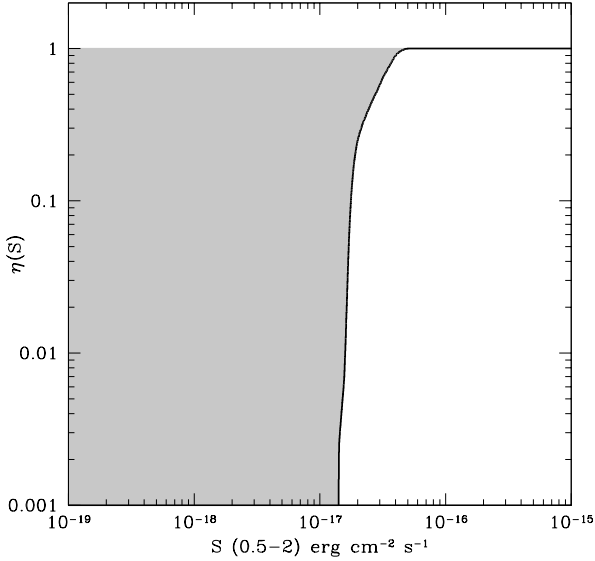}
   \caption{$Black~Line:$The selection function for point sources in the inner 5$\arcmin$ of the 4Ms CDFS.
   The $grey~shaded~area$ represents the complementary of the selection function (1-$\eta(S)$), highlighting the
   flux region from where the fluctuations are produced.}
              \label{skycov}
    \end{figure}
In Fig. \ref{skycov} we show the selection function, in  terms of fraction of field of view 
as a  function of the flux limits.  
If $\eta(S)$ is the selection function for sources with flux S, 
the source counts of sources producing the shot noise
is given by
\begin{equation}
\frac{dN_X}{dS}_{(SN)}=\alpha\,\left(1-\eta(S)\right)\,\frac{dN_X}{dS}.
\end{equation}
 $\alpha$ is a factor that accounts for the lower normalization observed in the 
CDFS with respect to larger area fields and with the \citet{gil07} model \citep{cap09}. We 
measured $\alpha$ by comparing the \citet{lehmer12} 0.5-2 keV logN-logS
and the \citet{gil07} modeled logN-logS at the break flux and obtained
$\alpha$=0.72.
We fixed the level of shot noise in our data by computing the 
integral in Eq. \ref{sn}, by assuming that the only sources accounting
for the SN are AGN and Galaxies. Therefore in this case we adopt:
\begin{equation}
\frac{dN_X}{dS}_{SN}=\frac{dN_X}{dS}_{AGN}+\frac{dN_X}{dS}_{Gal}.
\end{equation}
 dN/dS$_{AGN}$ has been estimated
from the population synthesis model of \citet{gil07} and   dN/dS$_{Gal}$
has been computed with the model of X-ray Galaxy evolution described in 
Sect. \ref{gal}. 
 In our calculations fluxes have been converted into count-rates by assuming an average power-law spectrum 
 with spectral index $\Gamma$=1.7. With such an assumption count-rates are obtained by multiplying
the flux by  an energy conversion factor ECF= 1.66$\times$10$^{11}$ cm$^{2}$/erg. Note that a different choice of the
 spectral index of $\Delta\Gamma\sim\pm{0.3}$ produces a 5\% variation of the  ECF.
The shot noise PS is scale independent (white noise) and since the signal from  cosmological sources
 increases with the scale, the contribution of P$_{2,SN}$ to the total PS becomes more important at small angular scales. \\
According to these prescriptions, we obtained an estimate of P$_{SN}$=1.58$\times$10$^{-12}$ \psu. 
The shot noise, as well as the other astronomical components of the PS, are affected by instrument PSF.
This effect consists in a multiplicative factor that applies to the effective PS. 
In order to model it we used the empirical approach proposed by \citet{chura} where
 the PS of each modeled component is multiplied  at every frequency by  factor:
\begin{equation}
P_{PSF}(k)=\frac{1}{[1+(k/0.12)^2]^{1.1}},
\end{equation}
where $k$ is the angular frequency.
In Fig. \ref{modelps} we show  our estimate of   P$_{SN}$,  taking into account the effects of the PSF.
When compared with the data we find an excellent agreement of our estimated   P$_{SN}$
with the Cosmic PS at short scale  (i.e. $\theta<$20$\arcsec$) and we interpret
this as a confirmation of our assumptions.   
Note that after subtracting the P$_{SN}$ from P$_{2,tot}(k)$, the remaining 
signal can be completely attributed to clustered sources and therefore
to the sources contributing to the unresolved extragalactic CXB. 
In order to evaluate the strength of our detection, we computed the probability
that our data could be obtained with a random fluctuation of 
a null flat signal. If we subtract the SN component,  at $\theta>$50$\arcsec$, 
we still observe a strong signal. In order to determine the significance of the 
remaining signal, we  fitted our data, on the whole angular range, with a P$(k)$=0 model (zero power signal) to determine 
if our signal could be due to a statistical fluctuation of a null signal.
With such a  model we obtained  $\chi^2$/d.o.f.=43.5/13,
thus rejecting our assumption  at $>$4$\sigma$ confidence level. 
Since such a signal increases with  scale, we interpret this as clustered cosmic signal.
\subsection{Anisotropies from AGN clustering} 
\begin{figure}
   \centering
 \includegraphics[width=0.5\textwidth, angle=0]{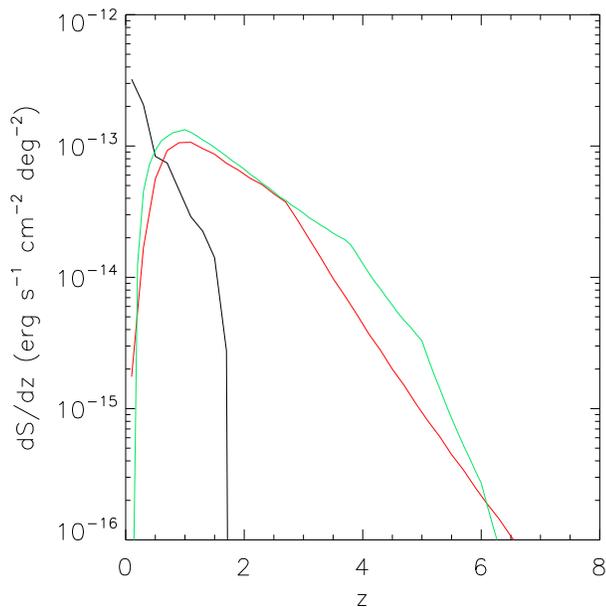}
   \caption{CXB production rate for undetected AGN ($Red~continuos~line$).
    Our prediction for the undetected galaxies CXB production rate is shown as $Green~Line$. 
    The $black~line$ represents the 
    emission from the IGM.}
              \label{agnflux}
    \end{figure}

On the small angles sampled by our data, P$_{2,CXB}$ is related to the unresolved CXB production 
rate d$S$/d$z$, and the evolving 3D PS of the AGN, P$_{3,AGN}(k)$ via the Limber's equation \citep{pee80}. 
At any given redshift the PS of AGN can be related to the PS of matter by knowing their redshift dependent linear biasing
factor \citep[b(z),][]{kai}.  Basically we have:
\begin{equation}
P_{3,AGN}(k,z)=b(z)^2 P_{3,M}(k,z),
\end{equation}
where  P$_{3,M}(k,z)$, is the matter 3D PS. 
 P$_{3,M}(k,z)$ was estimated by using the  CAMB\footnote{http://camb.info/} \citep{lewis} tool
 and including in the computation both the linear and the non-linear components of the 
 matter PS \citep{smith}.  Although a large amount of work  on galaxies and AGN clustering 
 is present in the  literature, here we choose to combine this cosmological tool
 with bias measurements and predictions because of the poor sampling of the 
 redshift space of observations. 
 The 2D PS can be then obtained with: 
\begin{equation}\begin{split}
&P_{2,AGN}(k)=\int^{z}_{0}\left(\alpha\,\frac{dS}{dz}\right)^2_{AGN}\times\\
&\frac{P_{3,AGN}(k[2\pi\,d_A*(1+z)]^{-1},z)}{c~dt/dz~[d_A*(1+z)]^2}~\frac{dz}{1+z},
\label{PS}
\end{split}\end{equation}
where d$_A$ is the angular diameter distance, and the integration is performed in the redshift range 0$<z<$7.5.  
A critical point of the analysis is the determination of the flux produced by undetected AGN at any 
given redshift.  We derived $\frac{dS}{dz}$ by using the  CXB synthesis model published by \cite{gil07}.
Briefly, the models take into account the observed  luminosity function, 
$k$-corrections, absorption distribution and spectral shapes  \citep[see][for more details]{gil07}
 of AGN and returns
the observed flux at any given redshift by using:
\begin{equation}
\frac{dS}{dz}=\int^{\infty}_{0}\,\left(1-\eta(S)\right)\int^{z+dz}_{z}\frac{L^\prime}{4\pi d_{L}^{2}}\,\phi(L^\prime,z)\,\frac{dV}{dz}\,dL^\prime \,dz
\label{prrate},
\end{equation}
where  d$_{L}$ is the luminosity distance, L$^\prime$ is the luminosity  measured
  in the 0.5(1+z)-2(1+z) keV range
and  dV/dz is the  comoving volume element.  The values of  $\frac{dS}{dz}$ have been calculated 
with the interactive tool provided by \citet{gil07}\footnote{http://www.bo.astro.it/$\sim$gilli/counts.html}.
The population of AGN used to compute the 0.5-2 keV CXB production rate has the following properties: 
42$<Log(L_X)<$47 erg/s, 20$<$Log$(N_{H})<$26 cm$^{-2}$, and
0$<z<$7.5. In addition, we included a high-redshift decline of the AGN space density \citep[see e.g.][]{bru09,civ11}. 
They indeed modeled the evolution of AGN X-ray luminosity function (XLF) at z$>$2.7 with an exponential decay 
($\phi(L,z)$=$\phi$(L,z$_0$)$\times$10$^{-0.43(z-z_0)}$ and z$_0$=2.7)
on top of the expected extrapolation from lower redshift parametrization of \citet{has05}. 
In Fig. \ref{agnflux} we show  the undetected AGN CXB production rate as function of the redshift.
The bulk of their flux  contributing to the unresolved CXB comes from  z$\sim$1.

\begin{figure}
   \centering
 \includegraphics[width=0.45\textwidth]{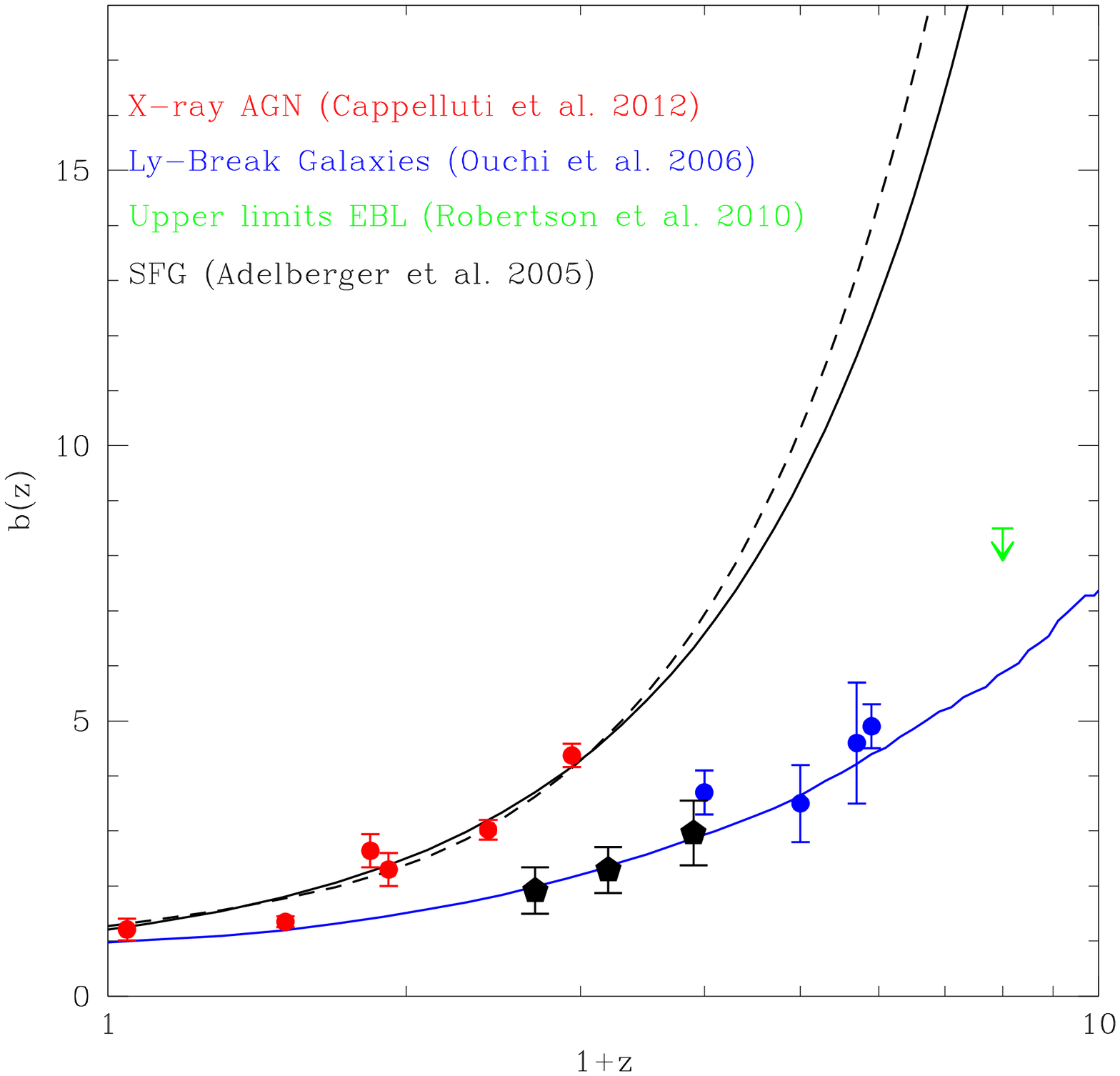}
   \caption{$Black~Continuous~Line$: Bias evolution model adopted for AGN compared with data measured by \citet{alle}; ($red~dots$).
   $Black~dashed~line$:  Bias evolution model for AGN of \citet{bon08}. $Blue~line$:  Bias
evolution    model assumed for X-ray emitting galaxies. $Black~pentagons:$ Bias measurements for star-forming galaxies from
\citet{ade} and with $blue~circles$ the measurements of bias of high-z Ly-Break galaxies of \citet{ouchi04}. In $green$ the upper-limit
from EBL fluctuations of Robertson et al. (2010)}
              \label{bias}
    \end{figure}

\begin{figure*}
   \centering
 \includegraphics[width=0.5\textwidth,angle=90]{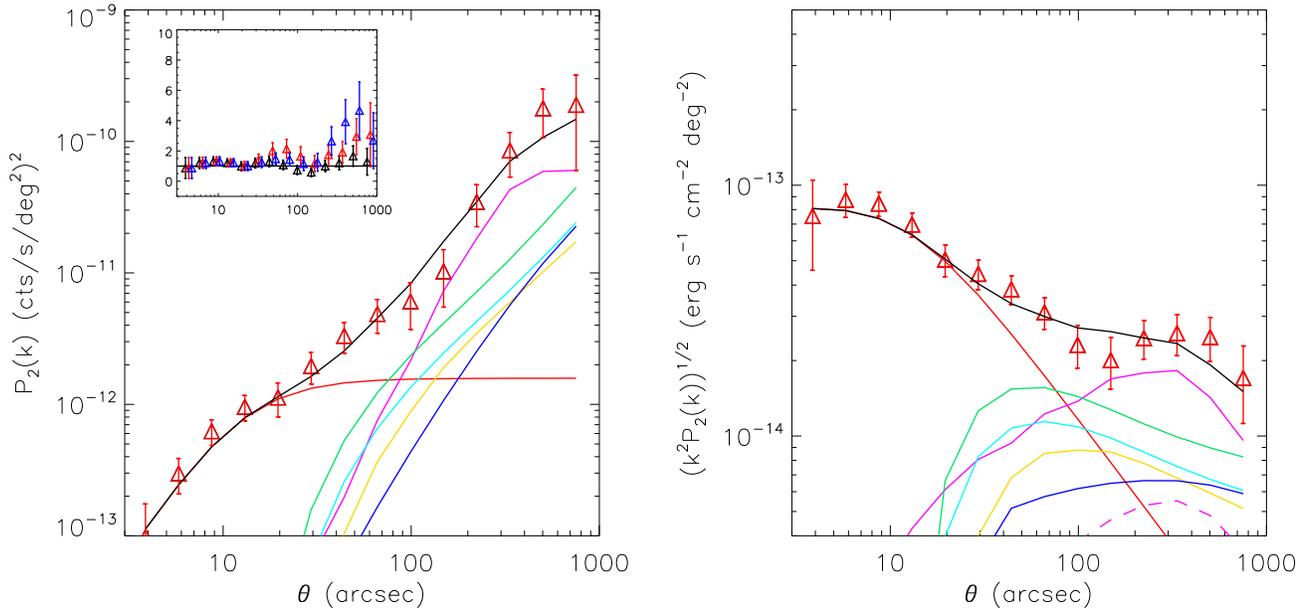}
   \caption{Left Panel. $Red~Triangles:$ Cosmic PS in the CDFS as a function of the scale. $Red~continuous~line:$ PS of the shot noise 
   (Galaxies and AGN).
$Green~continuous~line:$ PS of  undetected AGN clustering. $Cyan~line$: 
PS of undetected galaxies clustering. $Magenta~continuous~line:$ Total power
 from undetected hot gas in the  IGM. $Blue-line:$ predicted clustering  from of mini-quasars. 
 The $black~line$ represent to overall model described in Eq. \ref{hyp}.
 The inset shows the data/model ratio as a function of the scale for a SN and AGN+Galaxy clustering model ($Blue$),
 SN+IGM clustering model ($Red$) and SN+ AGN, Galaxies and IGM  clustering model ($black$).
 $Right~Panel$: Amplitude of the fluctuations as a function of the angular scale with the
  same color coding as in the left panel    \label{modelps} with the addition of 
 the contribution of
 IGM with 10$^{5}<T<$10$^{7}$ K and overdensity $\delta<$1000 following 
 the classical definition of WHIM  $Magenta~dashed~line$
 to enhance its contribution to the total IGM flux. $Yellow~line:$ Cross-correlation contribution.}
               \end{figure*}

Another important ingredient in the computation of P$_{2,AGN}(k)$ is the bias evolution of X-ray selected AGN.  
It has recently become clear that below z$\sim$3 AGN bias evolves like  that of DM halos (DMH) of mass
of $\sim$10$^{13.1}$ M$_{\odot}$ \citep[see e.g.][for a review]{cap12}. 
However, at higher redshift the bias factor of X-ray AGN is still unknown, while for 
optically selected QSOs this has been modeled up to z$\sim$5   \citep{hop07,bon08,dangle} with quadratic polynomials.
In Fig. \ref{bias} we compare the prediction of \citet{bon08}, rescaled to fit the z=0 X-ray 
selected AGN bias,  and the bias computed from analytical 
models \citep{vdb02,She01} for DMH with mass $\sim$10$^{13.1}$ M$_{\odot}$. As one can notice, 
both curves fit in an excellent way the observational data.  
 Since the evolution at high-z of the AGN bias 
factor is unknown and still matter of debate, we  assumed that the AGN bias
evolves like the bias of DMH of mass 
$\sim$10$^{13.1}$ M$_{\odot}$ up to z$\sim$10.
The contribution of AGN to the overall PS is shown in Fig. \ref{modelps}.

\subsection{Anisotropies from X-ray galaxy clustering \label{gal}} 
For anisotropies from galaxies  we 
adopted a procedure similar to that used for AGN.
The CXB production rate of galaxies\footnote{We considered "galaxies" all those X-ray 
sources with L$_X$(0.5-2) keV$<$10$^{42}$ erg/s.} has been obtained 
by folding in Eq. \ref{prrate} the z=0 luminosity function 
measured by \citet{Piero06}  with the evolution measured 
for star forming galaxies by Bouwens et al. (2011). The latter provides
the most recent measurement of the evolution of star formation in the Universe up to z$\sim$10
obtained with HST-WFC3. 
X-rays from non-active galaxies are mostly produced by low and high-mass X-ray binaries.
 Such objects with X-ray luminosities of  $\gtrsim10^{37}$ erg/s cannot be detected individually in distant galaxies. 
 The contributions from fainter discrete sources (including cataclysmic variables, active binaries, young 
 stellar objects, and supernova remnants) are well correlated with the star formation rate of the galaxy itself.
However the ignition of X-ray activity in the stellar population of galaxies
has a delay from the burst in star formation of the order of the time scale 
of stellar evolution of the donor star in binary systems.  
 Our  model
does not take into account effects of the delayed switch on  of 
X-ray Binaries from the star formation as done  by e.g Ptak et al. (2001); 
however, our representation allowed us to model the evolution of X-ray 
emitting galaxies up to z=10 with  the most recent results on star formation 
evolution.\\
  As shown by  \citet{Piero06}, 
the X-ray spectrum of galaxies can be represented by a simple power-law with spectral 
index $\Gamma\sim$2. With such an approximation no $k$-correction is needed.  \\
In Fig. \ref{agnflux} we show the unresolved CXB production 
rate of galaxies as a function of the redshift.
The galaxies  contribution to 
the total flux of the unresolved CXB is dominant with respect to AGN at  0$<$z$<$6.5.
Another parameter, that enters into the determination of the contribution of galaxies to the 
X-ray PS, is their bias factor and its evolution.  Most of the modern 
galaxy clustering analysis papers make use of the Halo Occupation formalism 
and therefore it is not possible to derive an analytical formulation of the bias evolution. 
Moreover, galaxy clustering is complicated by the luminosity dependence 
of clustering \citep{ouchi04}.
We therefore adopted a simple approximation for its determination, by assuming that the comoving 
correlation length of star forming galaxies is  constant, in the redshift range 0$<$z$<$7.5, at the value measured by \citet{ade} of 
 r$_0$= 4.5 Mpc/h
and their spatial correlation function can be modeled with a power-law with $\gamma$=1.6. 
Within this  scenario the bias factor of galaxies can be estimated at every redshift via \citep{pee80}:
\begin{equation} 
b(z)=\sigma_{8,G}(z)/\sigma_{8,DM}(z),
\end{equation} 
where $\sigma_{8,G}$(z) is the rms fluctuations of the galaxies distribution 
over a sphere with radius of 8 Mpc/h and $\sigma_{8,DM}$(z) is the same quantity 
for DM normalized to  $\sigma_{8,DM}$(z=0)=0.83.  
For such a power-law formalism it is possible to demonstrate that 
\begin{equation} 
(\sigma_{8,G})^{2}=J_2(\gamma)\left(\frac{r_0}{8 Mpc/h}\right)^{\gamma},
\end{equation} 
where J$_2$=72/[(3-$\gamma$)(4-$\gamma$)(6-$\gamma$)2$^{\gamma}$] \citep{pee80}.
In Fig. \ref{bias} we compare our predicted bias evolution with the measurement of 
high-z Lyman break galaxies of \citet{ouchi04} and with upper limits derived by Robertson et al. (2010).
As expected, galaxies are much less biased than AGN. 

\subsection{Emission from Cosmological structures and WHIM}
In  the inset of Fig. \ref{modelps} we show that, after including the contribution of the components mentioned above, the power spectrum still 
shows a prominent excess signal which, although with large error bars, increases up to a factor$\sim$3-4 toward low frequencies.  
Another significant contribution to the total signal of the CXB may arise from emission and clustering of unresolved 
galaxy clusters, groups and filaments.  From the sensitivity maps of galaxy clusters we determined the flux
limits for galaxy cluster/group detection and converted into a luminosity limit at every redshift.    
For galaxy clusters and groups luminosity , mass and temperature are related by scaling relations. 
The relation adopted here are discussed in \citet{fin07}.   The luminosity limit can be then translated into a mass limit
at every redshift.
Thus our source detection ensures the removal 
of galaxy clusters and groups down to a Log(M)=12.5-13.5 M$_{\odot}$ (i.e. kT$<$1.5 keV). Thus
only the low luminosity (low mass) and warm population of galaxy groups contributes to the unresolved CXB. 
\begin{figure}
 \centering
\includegraphics[width=0.45\textwidth]{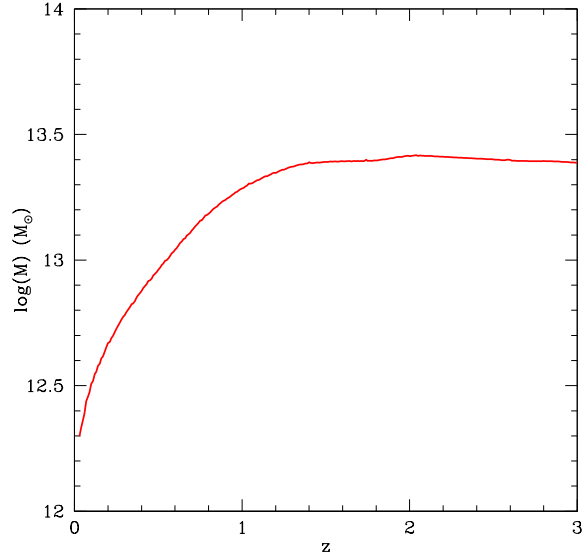}
\caption{ \label{mlim} The $Red~line$ represents the lowest detectable mass  in the 4Ms CDFS as a function of  redshift
 obtained through scaling relations.  }
             \end{figure}
Since this class of objects is very difficult to model analytically,  
we describe their properties using a set of mock maps from Roncarelli et al. (2012), who used a cosmological 
hydrodynamical simulation  to define the expected X-ray surface brightness due to the large scale structures (LSS).
The original hydrodynamical simulation (see the details in Tornatore et al. 2010) follows the evolution of a comoving 
volume of 37.5$h^{-1}$ Mpc$^3$ considering gravity, hydrodynamics, radiative cooling and a set of physical processes 
connected with the baryonic component, among which a chemical enrichment recipe that allows to follow the evolution of 7 
different metal species of the intergalactic medium (IGM).
 From its outputs, Roncarelli et al. (2012) simulated 20 lightcones, covering the redshift interval 
$0<z<1.5$ with a size of $\sim$0.25 deg$^2$ (roughly 4 times the size of the CDFs) each, and  angular resolution 
of 3.5 arcsec. Each pixel of the maps contains information about the expected observed spectrum in the 0.3--2.0 keV band with an 
energy resolution of 50 eV. The emission coming from the IGM was computed assuming an   emission  from an optically thin 
collisionally-ionized  gas ({\it Apec} in XSPEC)  model and considering the 
abundances of the different metal species provided by the simulation.
These maps/spectra have been convolved with the Chandra response
in order to  reproduce the effective Chandra count rates.  
Since our data are masked for galaxy clusters, we applied  to the simulations
a source masking   similar   to that on the real data.  
The unresolved CXB production rate evaluated from simulations is shown in Fig. \ref{agnflux}.
We have simulated   observations with the actual depth of the CDFS starting 
from the count rate maps described above, folded through a flat exposure map. We have added an artificial isotropic 
particle and cosmic background   according to the levels  estimated by \citet{HM06,HM07}. 
Random Poisson noise was artificially added to the image and we ran a simple sliding cell detection 
with a signal to noise ratio threshold of 4. We have then excluded all the regions within which  the overall 
encircled signal  from sources is above 4 sigma with respect to the background. 
Since galaxy clusters and groups are highly biased, in order to smooth out effects of sample variance
we extracted the power spectrum from all the masked maps  and averaged the results from all the 
realizations. Results of the PS modeling from undetected IGM is shown in Fig. \ref{modelps} (magenta solid lines).
A long standing debate in astrophysics is the possibility of detecting signal in emission or in absorption from the 
WHIM. In order to evaluate the contribution of WHIM to the overall PS signal 
of the unresolved CXB, we have extracted from our simulations  the same lightcones but
 including only all those photons with a temperature 10$^{5}<$kT$<$10$^{7}$ K
coming from regions with matter overdensity $\delta<$1000. Such a selection is compliant to the classical definition of the 
WHIM, even if some denser clumps might be present also inside the filamentary structures 
(see the discussion in Roncarelli et al. 2012). We have masked the WHIM photon maps 
 with the same masks used for clusters, extracted the PS  and averaged over the 20 realizations. 
The resulting modeled PS is shown in Fig. \ref{modelps} (magenta dashed lines).
\subsection{Cross-Correlation terms}
{As mentioned above, AGN, Galaxies and IGM partially 
share the same large scale structures. For this reason, fluctuations 
are boosted by their cross-correlation term. 
The 3-D cross-power spectrum (CPS) of the source populations $1$ and $2$
is determined by:
\begin{equation}
P_{3_{1,2}}(k)=b_1^2\,b_2^2\,P_{3,M}(k,z),
\end{equation}
where b$_1$ and b$_2$ are the bias factors relative to the source class 1 and 2,
respectively. 
In analogy with the PS, the angular CPS of diffuse emission 
produced by two populations with background production rate $\frac{dS}{dz}_1$ and 
$\frac{dS}{dz}_2$, respectively, can be evaluated with the following
form of the Limber's equation:

\begin{equation}\begin{split}
& P_{2_{1,2}}(k)=\int^{z}_{0} \alpha^2\left(\frac{dS}{dz}\right)_{1} \left(\frac{dS}{dz}\right)_{2} \\
& \frac{P_{3_{1,2}}(k[2\pi\,d_A*(1+z)]^{-1},z)}{c~dt/dz~[d_A*(1+z)]^2}~\frac{dz}{1+z}.
\label{CPS}
\end{split}
\end{equation}
We have then computed P$_{2_{AGN,IGM}}$, P$_{2_{AGN,Gal}}$ and P$_{2_{Gal,IGM}}$
using the values of bias and emissivity described above. As far as IGM is concerned
we adopted the bias evolution derived from simulations where b$_{IGM}(z)=\sqrt{1+z}$. 
The overall cross-power term can be then expressed as:
\begin{equation}
P_{2,A,G,I,}=P_{2_{AGN,IGM}}+ P_{2_{AGN,Gal}}+P_{2_{Gal,IGM}}.
\end{equation}
The cross power term is plotted in yellow in Fig. \ref{modelps}.
\subsection{Accuracy of the PS model}
\begin{figure*}
 \centering
 \includegraphics[width=0.45\textwidth]{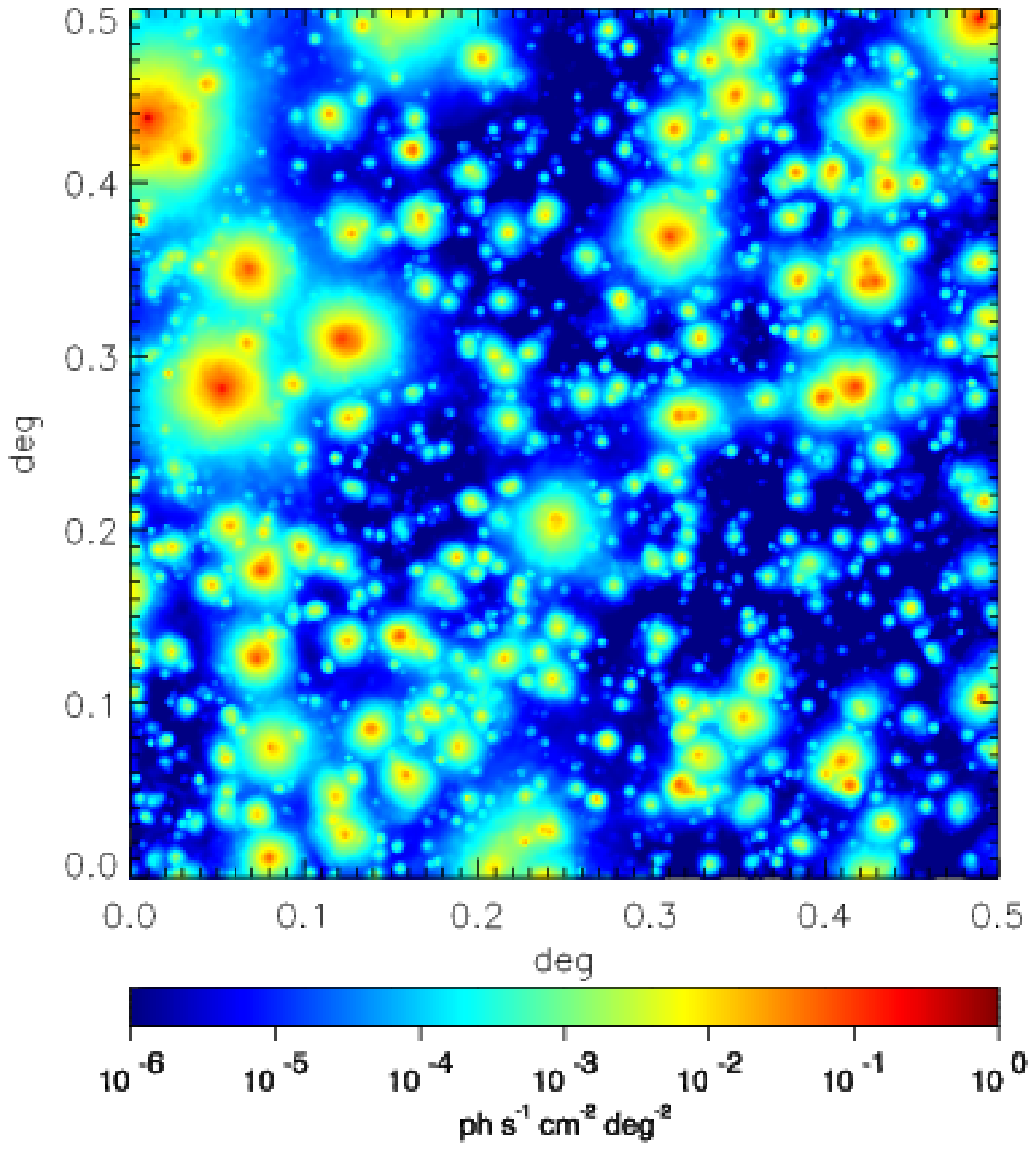}
\includegraphics[width=0.45\textwidth]{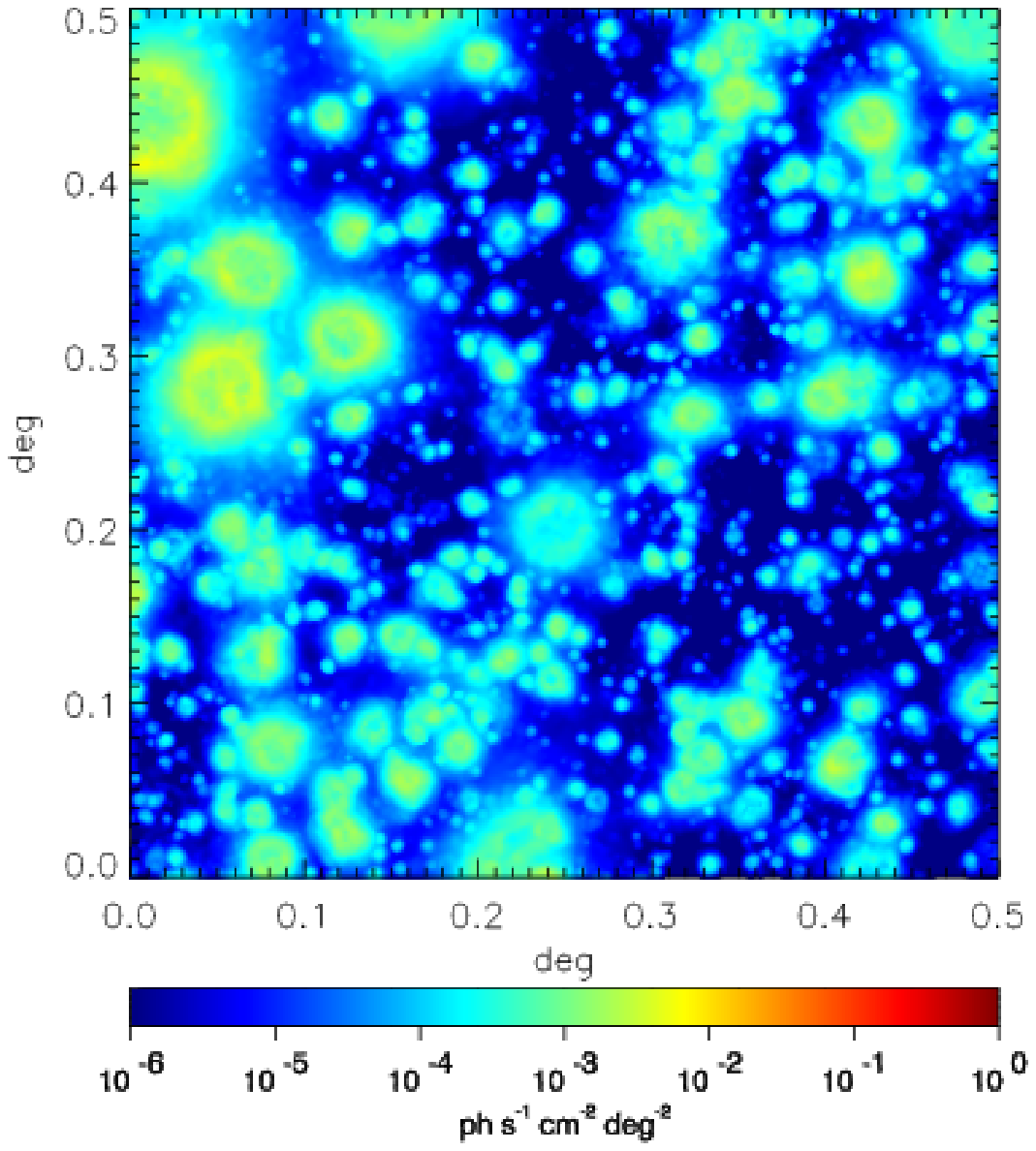}
  \caption{$Left~Panel:$ The whole simulation of the X-ray flux from Cosmological structure.  $Right$: 
  The same but for  IGM with 10$^{5}<$T$<$10${^7}$ K and  overdensity $\delta<$1000.  }
\end{figure*}

Our CXB PS  model is shown in Fig. \ref{modelps} where it is compared with data. 
Here we summarize the ingredients adopted to model the PS
components and their limits.
\begin{itemize} 
\item At small separations most of the signal is due to shot noise
	produced by AGN and galaxies within the {\em Chandra} 
	beam (red continuous line).
\item The AGN clustering is modeled by 
	convolving the matter PS for a concordance $\Lambda$-CDM
	cosmology, the AGN CXB production rate (dS/dz) derived from their X-ray luminosity function,
	and a  linear bias evolution of DMH of mass Log(M)=13.1 M$_{\odot}$. 
	Such a component  increases with the scale and 
	 its rms-fluctuations (right panel) peak on scales of the order of  40-50$\arcsec$.
	 Sources of uncertainties for such a component can be introduced by a) our 
	 assumption on the bias evolution that is unknown at z$>$2.5,  b) the 
	 evolution of their luminosity function has been determined at high-z 
	 only for most luminous sources and c) the assumption at the basis of the
	 \citet{gil07} population synthesis model. \citet{fiore} suggest that low-luminosity 
	 AGN may show a milder evolution at high-z with respect to observations 
	 at high luminosity. This may result in an underestimation of the power
	 at large scale. However, as discussed above, the peak of the 
	 undetected CXB flux is produced by AGN at z$\sim$1.5, where 
	 current observations are broadly consistent.  
\item X-ray galaxies clustering  (cyan line) is modeled  by 
	convolving the matter PS for a concordance $\Lambda$-CDM
	cosmology, a model of  the X-ray luminosity function obtained 
	by assuming an evolution similar to that of star forming galaxies
	starting from a measured z=0 XLF and a linear bias evolution 
	of sources having a correlation length r$_0$=4.5 Mpc/h.
	As for the AGN, such a component  increases with  scale
	and its  rms-fluctuations peak (right panel) on scales of the order of 50$\arcsec$.
	The sources of uncertainties introduced by our models are, like in the case
	of AGN, twofold and basically driven by our choice of the bias and XLF evolution.
	In the X-ray band the XLF of galaxies is known only up to z$\sim$1.5 where it 
	follows the evolution of star formation. 
	The limit of our approach is on the adopted model describing their 
	cosmological evolution,
	 which is however based on the reasonable assumption that X-ray galaxies 
	trace star formation in the Universe.
	The same argument applies to bias, where no
	measurements have been performed in the X-ray band.
	However, also in this case, observational proxies are consistent with our
	assumptions at the redshift where the bulk of the emission is produced,
	and therefore limiting the margin of uncertainty of our modeling. 
	Moreover, as we describe below, our modeling of the shot noise of AGN and galaxies
	is in good agreement with the data, meaning that our estimated source counts
	below the flux limit is statistically robust.
	\item The diffuse gas contribution (Magenta continuous line)  is modeled by computing the PS
	of maps obtained by hydrodynamical cosmological simulations  
	which include a  specific recipe of chemical enrichment.  
	Sources detectable in real observations have been excised from these 
	simulations before computing the PS.
	Such a component is responsible for most of the large scale fluctuations
	that peak on a few arcminutes scale.  For comparison, we also evaluated 
	the expected contribution of WHIM (magenta dashed line) 
	to the overall fluctuations and found that, on the largest
	angular scales, it contributes for up to  $\sim$1/3 of the total power. 
	The main source of uncertainty for this component is definitely
	 made by the actual metallicity of the IGM in the lowest density phase
	  which has  poor observational constrains. In addition,
	  the largest scale clustering in the simulations may be underestimated
	  because of the limited volume sampled in our cones.   
	\end{itemize}

In order to test our statistical hypothesis,  we performed 
 a $\chi^2$ test and evaluated the improvement of the fit
by adding one component after the other.  We considered 
every component (which have their amplitude and shape fixed by our assumptions) 
as if they were  a  parameter of the fit for the evaluation of the 
degrees of freedom.
The first test has been performed to evaluate if the observed fluctuations could
be explained by the shot noise component only. As discussed above, such an hypothesis
is rejected at $>$4$\sigma$ confidence level. 
Then we tested if, together with a shot-noise component, clustering of AGN and galaxies
(including their cross power) improved the fit. In this way we obtained $\chi^2$/d.o.f.=17.5/12.  
  We also  performed an F-test to evaluate the probability that the obtained $\chi^2$
could be obtained by a statistical fluctuation of the SN only model and obtained P(F-test)=1.2$\times$10$^{-3}$,
thus the inclusion of such a component significantly improves the fit.
 However, the relatively high $\chi^2$/d.o.f. value
suggests that additional or different components are required.\\
A visual inspection suggests that the strongest feature in the PS is produced by the IGM feature. 
Thus we fitted the data with SN+IGM model and obtained  $\chi^2$/d.o.f.=18.3/12
corresponding to P(F-test)=1.6$\times$10$^{-3}$  with respect to the SN only model.
  For such a model we show the data/model ratio
in red in the inset of Fig. \ref{modelps}. \\
By adding the point source clustering (and all the cross power terms) to the latter model we obtain
$\chi^2$/d.o.f.=7.8/11 and therefore the F-test, computed  to the SN+IGM model 
provides P(F-test)=2.7$\times$10$^{-3}$, thus providing a further $\sim$3$\sigma$ improvement of 
the fit.  If we compare this fit with the SN only model the F-test probility is P(F-test)=7.9$\times$10$^{-5}$,
thus providing a significant improvement of the overall fit quality. 
The overall data/model ratio  is shown in black in the inset of Fig. \ref{modelps}. \\
\begin{table*}
\centering
\begin{tabular}{cccc}
\hline
 Model & 	 $\chi^2$/d.o.f&. P(F-test) \\
              &               (1)        &    (2)\\
 \hline
 \hline
P$_{SN}$ & 43.4/13 & & \\
P$_{SN}$+P$_{Gal}$+P$_{AGN}$& 17.5/12& 1.2$\times$10$^{-3}$  \\
P$_{SN}$+P$_{IGM}$& 18.3/12& 1.6$\times$10$^{-3}$  \\
P$_{SN}$+P$_{Gal}$+P$_{AGN}$+P$_{IGM}$& 7.8/11& 2.7$\times$10$^{-3}$ *  \\
P$_{SN}$+P$_{Gal}$+P$_{AGN}$+P$_{IGM}$+P$_{mq}$ & 8.3/10& *  \\
\hline
\end{tabular}
\caption{\label{chisquare} (1) The   $\chi^2$/d.o.f.  value for model with different components include and (2), the null hypothesis
probabilty obtained by adding an additional component with respect to the model with n-1 d.o.f. *:P(F-test)=7.9$\times$10$^{-5}$
with respect to the SN only model.}
\end{table*}
 According to the values reported in the table,
 we can safely confirm that the unresolved CXB in the CDFS
  can be explained
 with random and clustered signal from undetected point sources (AGN and galaxies), 
 IGM clustering components plus a cross correlation term. 
  \section{Fluctuations from early black holes?}
Potential contribution to the unresolved CXB and its structure can also come from very high redshifts 
overlapping with epochs usually identified with first stars era. It is widely expected that fragmentation 
within the first collapsing protogalaxies was much less efficient so that first stars were significantly
 more massive and short-lived and could have left behind non-negligible abundance of accreting 
 black holes. Since, in addition, the first black holes could have also formed directly during the first stars era, 
 these populations may supply an additional, and potentially measurable,  contribution to the unresolved CXB and its fluctuations.

Spitzer-based studies have revealed significant levels of source-subtracted cosmic
 infrared background (CIB) fluctuations which were proposed to originate in the first stars era 
\citep{kash05}. Indeed the remaining known galaxy populations have been shown to produce significantly
 lower levels of the CIB fluctuations \citep{kari} and there is no correlation in the large-scale structures between the
  Spitzer CIB maps at 3.6 and 4.5 micron and HST/ACS likely pointing to the high-$z$ origin of the excess CIB 
  \citep{kash07}. The excess fluctuations have been confirmed in the AKARI based analysis extending to 2.4 micron 
\citep{matsumoto} and the signal has now been measured to extend to $\sim 1^\circ$ exceeding the power over
 the remaining normal galaxies by well over an order of magnitude \citep{kash}. The level of the fluctuations at 3.6 
 micron is $\delta F_{\rm CIB}\sim 0.07$ nW/m$^2$/sr and their energy spectrum appears to follow the Rayleigh-Jeans 
 law, $\delta F_{\rm CIB}\propto \lambda^{-3}$, between 2.4 and 4.5 micron. 
 If the excess CIB fluctuation arises at high $z$, the sources producing it would have 
 projected angular density of $\sim (1-3)$arcsec$^{-2}$ (Kashlinsky et al 2007).
Under the assumption that the high-z sources measured by \citet{kash12}
rapidly evolve into   miniquasars ( or early black holes), 
we estimated that, if the CIB/CXB flux ratio is constant along the cosmic time,
with  3.6 $\mu$m extragalactic CIB flux of  $\sim$6 nW m$^{-2}$ sr$^{-1}$ \citep{kash} and extragalactic CXB flux 
of 8.15$\times$10$^{-12}$ \cgs~deg$^{-2}$ \citep{lehmer12}, we obtain $\frac{S_{CIB}}{S_{CXB}}\sim$220.
Thus we evaluated the  contribution to the CXB  of the CIB fluctuations to
 be of the order   $\sim$3$\times$10$^{-13}$/(1+z) \cgs~deg$^{-2}$ (i.e. $\sim$0.5\%
of the CXB if these sources have z$\sim$7.5-15)\footnote{Assuming b(z)=$\sqrt{1+z}$}.  
We have therefore predicted the expected PS of these sources 
by folding their CXB production rate in Eq. \ref{PS}
and integrating in the redshift range 7.5-15.\\ 
The result of  this prediction is shown with a blue line in Fig. \ref{modelps}. 
A $\chi^{2}$ fit of our 5 components model provides $\chi^2$/d.o.f=8.3/10.
Although the inclusion of such a component does not improve the quality of the 
fit, the statistics does not allow us to  reject the hypothesis that these sources
contribute to the observed fluctuations. 
These sources would contribute very weakly to the shot noise,
and the upper limit of their contribution is not larger than the uncertainty on the measured PS. 
Thus at 1$\sigma$ we have P$_{SN, mq}\lesssim$1$\times$10$^{-13}$\psu (i.e. the mean value of the uncertainty
on those scales). According to Eq.\ref{sn},
we can estimate the source density of putative miniquasars with P$_{SN}$/S$_{min}^{2}\sim$N($>S_{min}$). 
If S$_{min}$=10$^{-20}$\cgs~, where most CXB models predict the saturation of the source counts,
 we estimate N$\lesssim$60.000 deg$^{-2}$. If these sources shine
in the redshift range 7.5-15 \citep{kash12}, then their comoving volume density is $\le$9$\times$10$^{-5}$ Mpc$^{-3}$.
For comparison in the same redshift range the \citet{gil07} model predicts, for X-ray selected AGN, in the luminosity 
range sampled by our fluctuations (Log(L$_X$)$\lesssim$43.58 erg/s, at z=7.5-15, for sources
with observed -20$\le$LogS(0.5-2)$\le$-17 \cgs), source densities of the order   1.5$\times$10$^{-4}$ Mpc$^{-3}$ and
8.4$\times$10$^{-5}$ Mpc$^{-3}$ in the case of flat and declining evolution, respectively. 
The upper limit derived above  suggests that the declining evolution of AGN
is a good representation of the evolutionary track of these remote sources. 
\section{Contribution of undetected source populations to the CXB}
By interpreting the observed 
behavior of the unresolved CXB fluctuations, 
we have developed a model which 
is able to explain  the 
nature of the unresolved CXB.
The fluctuations observed here are 
reproduced in the PS if the clustering recipes
described in the text are combined with the 
unresolved CXB production rates
shown in Table \ref{dsdz}.
\begin{table*}
\centering
\caption{\label{dsdz} (1) CXB production rate of every 
class of undetected sources. (2) The percentage of
the overall and  (3) unresolved CXB produced by every population.
 In the last line we show the cumulative values, 
the ranges refer to Model I and II.} 
\begin{tabular}{cccc}
\hline
\hline
 Component  & 	 S & \% CXB$^{a}$ &  \% Unres. CXB \\
                         & $\times$10$^{-13}$\cgs~deg$^{-2}$ &                 \%                         &\%\\
                         & (1)  & (2) & (3) \\
\hline
 AGN    &          1.97             &       2.4$\pm{0.2}$   &       19.3$\pm{1.3}$       \\   
 Galaxies   &          2.51             &         3.1$\pm{0.2}$  &           24.6$\pm{1.7}$  \\
 IGM           &         5.70             &     7.0$\pm{0.5}$       &            55.9$\pm{3.9}$  \\
$^{b}$WHIM       &      1.70                &  2.1$\pm{0.1}$           &          16.7$\pm{1.2}$    \\
 Early-BH  &  $<$0.35                   &  $<$0.5            &  $<$3.4    \\
 \hline
 Total         &  10.18   &   12.4$\pm{0.9}$ & 100 \\
 \hline
 \hline
 \label{dsdz}
\end{tabular}

\footnotesize
$^{a}$: Computed using a total 0.5-2 keV CXB flux of 8.15$\pm{0.58}\times$10$^{-12}$\cgs~deg$^{-2}$ \citep{lehmer12}.\\
$^{b}$:The WHIM flux is included in the IGM flux.\\
\end{table*}
Our results show that, in the 0.5-2 keV band, the effective 
fraction of the unresolved extragalactic background 
is of the order of 12\% of the total.   In Table \ref{dsdz} we show the CXB flux that our model
predicts to explain the fluctuations together with the fractions
of total and unresolved CXB produced by every undetected source population.
As one can notice in Table \ref{dsdz}, the bulk 
 ($\sim$56\%) of the unresolved CXB 
flux is made by unresolved clusters, groups and the WHIM which
accounts, by itself, for $\sim$17\% of the unresolved flux. \\
For  point sources our model predicts
 that  AGN and galaxies 
contribute, all together,  for the remaining flux of the unresolved CXB,
with galaxies and AGN producing $\sim$25\% and $\sim$20\% of the unresolved flux, 
respectively. \\
Moreover, our data cannot exclude that a sizeable fraction of the unresolved
CXB could be produced by a population of still undetected
high-z sources, likely black hole seeds. 
 \section{Discussion}
 In this paper we presented the measurement of 
 the angular PS of the fluctuations of the unresolved CXB 
 in  the 4Ms observation of the CDFS in the angular 
 range $\lesssim10\arcmin$. 
Poisson noise and spurious signals have been modeled
and removed from the measured PS.  
We  performed a spectral decomposition analysis 
 and showed that after removing the 
  low frequency signal, which can be attributed to the 
 shot noise of unresolved sources that  randomly enter the beam,
 the amplitude of the fluctuations with extragalactic origin account for $\sim$12.3\% 
 of the CXB and the significance of the detection 
 of these cosmic fluctuations is  $>$10$\sigma$. 
 In the next section we briefly discuss to properties of the 
populations producing the unresolved CXB fluctuations.
 \subsection{The population of  undetected AGN}
 For AGN we folded the observed 
 evolution scenarios with the population synthesis model 
 of \citet{gil07}  and  a simple recipe for bias
 where AGN are tracing DMH with mass Log(M)=13.1 M$_{\odot}$. 

This  population of AGN 
has a  space density that exponentially declines above z=2.7. 
The CXB production rate necessary to produce the modeled 
PS yields to a fraction of the unresolved 0.5-2 keV CXB flux
of $\sim$19\%.   We predicted a CXB flux produced by undetected AGN of
$\sim$2.0$\times$10$^{-13}$ \cgs~deg$^{-2}$. 

We also tried to probe different evolution scenarios but 
our data do not allow to significantly constrain
the behavior of the AGN XLF at high-z.  
In Fig. \ref{lognlogs} we show the predicted LogN-LogS
that, according to our model, satisfies the 
observed fluctuations compared with recent observations. 

\subsection{The population of undetected X-ray galaxies}
 Galaxies are  the most numerous population 
of objects  contributing to the unresolved CXB (see Fig. \ref{lognlogs});
 the power produced by such a population 
is lower that  that  of AGN since  they are less biased 
even if they produce
 more CXB flux.
In the soft X-rays, galaxies have been observed up to z$\sim$1 \citep[see e.g.][]{lehmer}, 
and therefore their high-z space density is unknown.
We have developed a toy model for the galaxies XLF
where their evolution follows that of the star formation 
in the universe. 
 With this model we estimated that
 galaxies contribute to $\sim$25\% of the 
 unresolved  CXB flux.  In a recent paper, \citet{dij}
 used a modeling similar to ours  and found that,
 in principle, X-ray galaxies could produce all of the unresolved
 1-2  keV CXB. However  they did not 
 consider the contribution of other undetected sources that
 we have shown to produce a large fraction of the unresolved CXB.
 Overall, the predicted source counts of AGN and galaxies
 are in good agreement with the measurements of \citet{xue} and \citet{lehmer12} in the same field.
  \begin{figure}
 \centering
\includegraphics[width=0.45\textwidth]{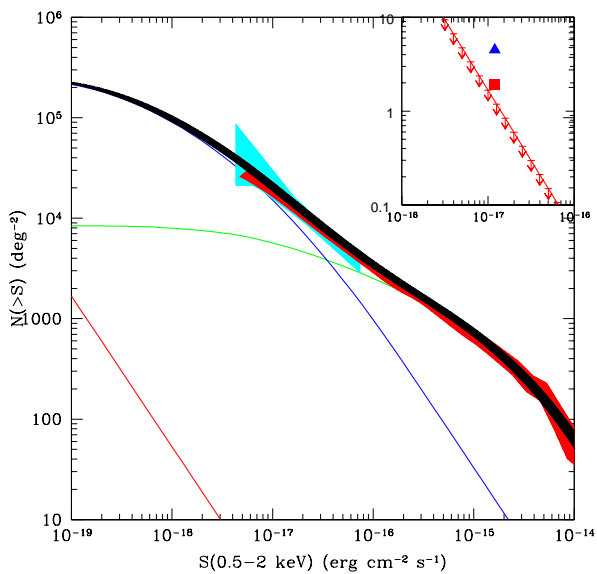}
\caption{\label{lognlogs}   The 0.5-2 keV logN-logS used to reproduce the observed 
unresolved CXB fluctuations. In $green$ AGN, $blue$ galaxies and, in  $black-shaded$, the sum of the AGN and galaxies,
the uncertainty is attributed to the count-rate to flux conversion.   The $red~shaded~area$ is the logN-logS measured by
\citet{lehmer12} in the CDFS. The $red~line$ 
represents the upper limit of the logN-logS of miniquasars. 
 The $cyan~shaded~area$ represents the expected counts from the fluctuation analysis of \citet{mi2}. 
In the inset we show a zoom onto the miniquasar upper-limit logN-logS. The $blue~triangle$ is the expected number of 
X-ray sources, at the flux limit of the 4Ms CDFS, produced at z$\lesssim$10 from collapse of  
POPIII stars accreting at the Eddington Limit. The $red~triangle$ is number of X-ray sources in case of 
of direct collapse (Quasi-stars) accreting at $\lambda_{Edd}$=0.3} 
\end{figure}

  \subsection{CXB from IGM and WHIM}
On scales larger than 100$\arcsec$ our analysis shows that the main contribution to  the unresolved 
CXB fluctuations is due  to  IGM emission  produced
by undetected groups, clusters  and the  WHIM.
This has been estimated with cosmological hydrodynamical simulations 
that, together with structure formation,  include
 feedback mechanisms that pollute the   IGM with metals
which are responsible for the X-ray emission  \citep{ron}. 
 Such a component produces $\sim$50\% of the unresolved
 CXB.\\
 We determined that 
$\sim$1/3 of the IGM flux is produced by the WHIM,
 where the so called "missing baryons"  are expected to lie.
 However the WHIM definition taken from simulations 
 already applies a density cut ($\delta < 1000$) that is 
 meant to mimic roughly the effect of excising detected sources. 
 In reality, dense clumps may be present outside virialized objects
  so we must consider our determination as a lower limit 
  of the total WHIM contribution (see the discussions in Roncarelli et al. 2006, 2012). 

Our model thus predicts that 
the  flux of the WHIM in the CDFS is of the order
1.7$\times$10$^{-13}$ \cgs~deg$^{-2}$ (i.e. 2.3\% of the 
total CXB flux) and  produces a signal peaking 
on a few arcmin scale. Such an estimate is about one order of magnitude lower than  what
measured by \citet{gale}  (i.e. 12\% of the overall diffuse emission in shallower 0.2-0.4 keV
XMM-{\em Newton} observations) where they did not model
the contamination from undetected sources.  
However such an estimate relies on the output of the 
simulation and is very sensitive to the metallicity of the WHIM.
According to \citet{ron},  different recipes for the metal enrichment of the 
WHIM may lead to a variation  up to a factor 3 in overall emissivity 
of the WHIM. More information on such a component of the Universe
will be possible if, for example, contamination of undetected cluster 
could be excised from X-ray maps by masking also optically detected
groups with mass Log(M)$>$12-12.5 M$_{\odot}$. 
\subsection{Very high-z sources}
 Finally, we speculated on   the  possible existence of 
 a population of high-z miniquasars (or early black holes) born from  the collapse
 of early massive objects. 
 Although we did not detect
 their signature, we placed an upper limit to their contribution to 
 the CXB  ($<$3$\times$10$^{-13}$/(1+z) \cgs~deg$^{-2}$, z$>$7.5).
 We estimated that 
 these sources would follow the declining  evolutionary track
 of AGN with Log(L$_X$)$\lesssim$43.58 erg/s.\\
 Our observations are not sensitive to the  faint fluxes expected from these sources,
 and thus we could only place upper limits.
However, since such a population peaks at z$>7.5$, their
fluctuations should peak on scales of the order of several tens of arcminutes, where 
the contribution from shot-noise is relatively weak also 
in shallower surveys. On such a scale,  foreground
source population PS significantly  dims and therefore
their detection would be possible. 
To conclude, we determined that at fluxes of the order 10$^{-20}$ \cgs, 
their number density is order of $\sim$60000 deg$^{-2}$ which means that,
under the assumption of Euclidean logN-logS,  at the flux limit of the CDFS
their number density is of the order 1-2 deg$^{-2}$. \\
Using the theoretical  predictions of \citet{volo}, \citet{treis} computed the expected
number density of the first X-ray sources at z$\sim$7-10 at the 4Ms CDFS flux limit.
 In the case of Population III stars remnants accreting at the Eddington limit
 they find 4.5 deg$^{-2}$, while in the case of direct collapse \citep[quasi-stars,][]{beg} 
 with accretion regime at $\lambda_{Edd}$=0.3 their expected 
 number density is  1.9 deg$^{-2}$. In Fig.  \ref{lognlogs} we show 
 a comparison of these predictions with our upper-limit which, at the zero-th order, is
 favoring the direct collapse scenario. \\

Our  work suggests that future deeper observations 
on wider fields would allow us to improve the sensitivity
of the PS measurement by reducing the Shot-Noise and 
masking fainter clusters. Moreover a deep survey 
with a flux limit comparable or deeper to that of the 4Ms CDFS
covering  1-2 deg$^{2}$,  would allow us a direct detection of the 
 WHIM feature and, by measuring the large scale shape 
 of the PS, investigate the very  high-z X-ray Universe. 
  A possible future X-ray mission like the proposed 
 Wide Field X-ray Telescope \citep[WFXT][]{wfxt} will be able to study the
 PS of the unresolved CXB with high precision given the large collecting
 area (low photon noise) and the large field of view (low cosmic variance).

\section*{Acknowledgments}
 NC acknowledges the INAF-Fellowship program for support. 
 PA acknowledges support from Fondecyt 11100449.
We acknowledge financial contribution from the agreement ASI-INAF  1/009/10/0.
FN acknowledges support from  NAS XMM Grant NNX08AX51G, XMM Grant NNX09AQ05G
and ASI Grant ASI-ADAE 
 PR acknowledges a grant from the Greek General Secretariat of Research and 
Technology in the framework of the programme Support of Postdoctoral 
Researchers. NC thanks the anonymous referee for the suggested improvements.

%
%
%

\def\aj{AJ}%
\def\araa{ARA\&A}%
\def\apj{ApJ}%
\def\apjl{ApJ}%
\def\apjs{ApJS}%
\def\ao{Appl.~Opt.}%
\def\apss{Ap\&SS}%
\def\aap{A\&A}%
\def\aapr{A\&A~Rev.}%
\def\aaps{A\&AS}%
\def\azh{AZh}%
\def\baas{BAAS}%
\def\jrasc{JRASC}%
\def\memras{MmRAS}%
\def\mnras{MNRAS}%
\def\pra{Phys.~Rev.~A}%
\def\prb{Phys.~Rev.~B}%
\def\prc{Phys.~Rev.~C}%
\def\prd{Phys.~Rev.~D}%
\def\pre{Phys.~Rev.~E}%
\def\prl{Phys.~Rev.~Lett.}%
\def\pasp{PASP}%
\def\pasj{PASJ}%
\def\qjras{QJRAS}%
\def\skytel{S\&T}%
\def\solphys{Sol.~Phys.}%
\def\sovast{Soviet~Ast.}%
\def\ssr{Space~Sci.~Rev.}%
\def\zap{ZAp}%
\def\nat{Nature}%
\def\iaucirc{IAU~Circ.}%
\def\aplett{Astrophys.~Lett.}%
\def\apspr{Astrophys.~Space~Phys.~Res.}%
\def\bain{Bull.~Astron.~Inst.~Netherlands}%
\def\fcp{Fund.~Cosmic~Phys.}%
\def\gca{Geochim.~Cosmochim.~Acta}%
\def\grl{Geophys.~Res.~Lett.}%
\def\jcp{J.~Chem.~Phys.}%
\def\jgr{J.~Geophys.~Res.}%
\def\jqsrt{J.~Quant.~Spec.~Radiat.~Transf.}%
\def\memsai{Mem.~Soc.~Astron.~Italiana}%
\def\nphysa{Nucl.~Phys.~A}%
\def\physrep{Phys.~Rep.}%
\def\physscr{Phys.~Scr}%
\def\planss{Planet.~Space~Sci.}%
\def\procspie{Proc.~SPIE}%
\let\astap=\aap
\let\apjlett=\apjl
\let\apjsupp=\apjs
\let\applopt=\ao

\label{lastpage}

\end{document}